\documentstyle[12pt,graphicx,subfigure]{article}                
\begin{document}
\baselineskip 18pt
\def\today{\ifcase\month\or
 January\or February\or March\or April\or May\or June\or
 July\or August\or September\or October\or November\or December\fi
 \space\number\day, \number\year}
\def\thebibliography#1{\section*{References\markboth
 {References}{References}}\list
 {[\arabic{enumi}]}{\settowidth\labelwidth{[#1]}
 \leftmargin\labelwidth
 \advance\leftmargin\labelsep
 \usecounter{enumi}}
 \def\newblock{\hskip .11em plus .33em minus .07em}

 \sloppy
 \sfcode`\.=1000\relax}
\let\endthebibliography=\endlist
%
\def\Journal#1#2#3#4{{#1} {~ #2}, #3 (#4)}

\def\NCA{\em Nuovo Cimento}
\def\NIM{\em Nucl. Instrum. Methods}
\def\NIMA{{\em Nucl. Instrum. Methods} A}
\def\NPA{{\em Nucl. Phys.} A}
\def\NPB{{\em Nucl. Phys.} B}
\def\PLB{{\em Phys. Lett.}  B}
\def\PRL{\em Phys. Rev. Lett.}
\def\PRD{{\em Phys. Rev.} D}
\def\ZPC{{\em Z. Phys.} C}
\def\lsim{\ ^<\llap{$_\sim$}\ }
\def\gsim{\ ^>\llap{$_\sim$}\ }
\def\r2{\sqrt 2}
\def\beq{\begin{equation}}
\def\eeq{\end{equation}}
\def\beqn{\begin{eqnarray}}
\def\eeqn{\end{eqnarray}}
\def\rmuu{\gamma^{\mu}}
\def\rmud{\gamma_{\mu}}
\def\PL{{1-\gamma_5\over 2}}
\def\PR{{1+\gamma_5\over 2}}
\def\sinW2{\sin^2\theta_W}
\def\AEM{\alpha_{EM}}
\def\mul{M_{\tilde{u} L}^2}
\def\mur{M_{\tilde{u} R}^2}
\def\mdl{M_{\tilde{d} L}^2}
\def\mdr{M_{\tilde{d} R}^2}
\def\mz2{M_{z}^2}
\def\c2b{\cos 2\beta}
\def\au{A_u}
\def\ad{A_d}
\def\cob{\cot \beta}
\def\v#1{v_#1}
\def\tb{\tan\beta}
\def\epem{$e^+e^-$}
\def\KK{$K^0$-$\bar{K^0}$}
\def\wi{\omega_i}
\def\xj{\chi_j}
\def\Wmu{W_\mu}
\def\Wnu{W_\nu}
\def\m#1{{\tilde m}_#1}
\def\mH{m_H}
\def\mw#1{{\tilde m}_{\omega #1}}
\def\mx#1{{\tilde m}_{\chi^{0}_#1}}
\def\mc#1{{\tilde m}_{\chi^{+}_#1}}
\def\mwi{{\tilde m}_{\omega i}}
\def\mxi{{\tilde m}_{\chi^{0}_i}}
\def\mci{{\tilde m}_{\chi^{+}_i}}
\def\mz{M_z}
\def\sw{\sin\theta_W}
\def\cw{\cos\theta_W}
\def\cb{\cos\beta}
\def\sb{\sin\beta}
\def\rwi{r_{\omega i}}
\def\rxj{r_{\chi j}}
\def\rfp{r_f'}
\def\Kik{K_{ik}}
\def\Fq2{F_{2}(q^2)}
\def\mg{m_{\frac{1}{2}}}
\def\mchi1{m_{\chi}}
\def\tw{\tan\theta_W}
\def\sec2w{sec^2\theta_W}

\begin{center}{\LARGE\bf WMAP Data and 
Recent Developments in Supersymmetric Dark Matter}\\
\vskip.25in
{Utpal Chattopadhyay\footnote{E-mail: tpuc@iacs.res.in}$^{(a)}$, 
Achille Corsetti\footnote{E-mail: corsetti@neu.edu}$^{(b)}$ and  
Pran Nath\footnote{E-mail: nath@neu.edu}$^{(b)}$  }

{\it
(a) Department of Theoretical Physics, Indian Association for the Cultivation 
of Science, Jadavpur, Kolkata 700032, India\\
(b) Department of Physics, Northeastern University, Boston, MA 02115-5005, USA\\
}
\end{center}

\begin{abstract}  
A brief review is given of  the recent developments in the 
analyses of supersymmetric dark matter. Chief among these is the
very accurate  determination of the amount of cold dark matter in 
the universe from analyses using WMAP data. The implications of
this data for the mSUGRA parameter space are analyzed. It is shown
that the data admits solutions on the hyperbolic branch (HB) of the
radiative breaking of the electroweak symmetry. A part of the hyperbolic
branch lies in the so called inversion region where the LSP neutralino 
$\chi_1^0$ becomes essentially a  pure Higgsino and degenerate with the
 next to  the lightest neutralino $\chi_2^0$ 
 and the light chargino $\chi_1^{\pm}$. 
 Thus some of the conventional signals for the observation of 
 supersymmetry at colliders (e.g., the
missing energy signals) do not operate in this region. 
On the other hand the inversion region contains a high degree of 
degeneracy of $\chi_1^0, \chi_2^0,\chi_1^{\pm}$ leading to 
coannihilations which allow for the satisfaction of the WMAP 
relic density constraints deep on the hyperbolic branch. Further,
an analysis of the neutralino-proton cross sections in this region
reveals that this region can still be accessible to dark matter 
experiments in the future. Constraints from $g_{\mu}-2$ and 
from $B^0_s\rightarrow \mu^+\mu^-$ are discussed. Future
prospects are also discussed.
\end{abstract}

\section{Introduction}
Very recently the data from the Wilkinson Microwave Anisotropy Probe 
(WMAP) has allowed analyses of the cosmological parameters to a high degree
of  accuracy\cite{bennett,spergel}. These analyses also indicate
unambiguously the existence of cold dark matter (CDM) and put sharp
limits on it. At the same time over the past decade experiments
for the direct detection of dark matter have made enormous 
progress\cite{dama,cdms,hdms,edelweiss} with reliable limits 
emerging on the CDM component in direct laboratory experiments.
Further, experiments are planned which in the future will be able
to improve the sensitivities by several orders of 
magnitude\cite{genius,cline,Smith:2002af}. 
In this talk we will give a brief review of the recent developments in 
supersymmetric dark matter (For a sample of other recent reviews see
Ref.\cite{darkreviews}). We will review the  constraints
on the analyses of dark matter from $g_{\mu}-2$ and from 
$B^0_{s,d}\rightarrow \mu^+\mu^-$. 
 We will also discuss the effects  of nonuniversalities
and the effects of the constraints of Yukawa coupling unification.
 One of the main focus of our analysis
will be the study of dark matter on the 
hyperbolic branch\cite{Chan:1997bi} (and focus point region\cite{fmm} which
is a subpiece of the hyperbolic branch) and 
its implications for the discovery of supersymmetry\cite{Chattopadhyay:2003xi}.
As is well known SUGRA models with R parity  provide a candidate
for supersymmetric dark matter. This is so because in SUGRA
unified models\cite{msugra, Nath:2003zs} one  finds that over a 
large part of the parameter
space the lightest supersymmetric particle (LSP) is the lightest
neutralino which with R parity conservation becomes a candidate
for cold dark matter (CDM). (An interesting alternate possibility
discussed recently is that of axionic dark matter\cite{Covi:1999ty}).
In the simplest version of SUGRA models\cite{msugra,Nath:2003zs}, 
mSUGRA, which is based on 
 a flat K\"ahler potential the soft sector of the theory
is parameterized by  $m_0, m_{\frac{1}{2}}, A_0, \tan\beta$,
where $m_0$ is the universal scalar mass, $m_{\frac{1}{2}}$ is 
the universal gaugino mass,  $A_0$ is universal trilinear
coupling and $ \tan\beta =<H_2>/<H_1>$ where $H_2$ gives mass
to the up quark and $H_1$ gives mass to the down quark and the
lepton. The minimal model  can be  extended by considering a 
curved K\"ahler manifold and also a curved  gauge kinetic energy
function. Specifically these allow one to include nonuniversalities
in the Higgs sector, in the third  generation sector  and in
the gaugino sector consistent with flavor 
changing neutral 
currents\cite{Nath:1997qm,nonuni1,Corsetti:2000yq,nelson,Chattopadhyay:2003yk}.

\section{Constraints on dark matter analyses}
There are a number of constraints that must be imposed 
in the theoretcal analyses of supersymmetric dark matter.
These include the constraints from $g_{\mu}-2$, the flavor
changing neutral current (FCNC) constraints, constraints from
$B^0_s\rightarrow \mu^+\mu^-$ limits, and constraints from
the currents limits on the relic density for cold dark matter.
Here we discuss some of these. We begin with a discussion of
the $g_{\mu}-2$ constraint.
The analysis of $g_{\mu}-2$ has been under scrutiny for a 
considerable  period of  time. In supersymmetry it is 
predicted that the effects of supersymmetric electroweak  corrections
to $g_{\mu}-2$ can be of the same size as the standard model 
electroweak effects\cite{Yuan:ww}. 
Furthermore, the sign of the supersymmetric electroweak correction
is correlated  with the sign of the Higgs mixing parameter 
$\mu$\cite{lopez,Chattopadhyay:1995ae}.
It is also known that the effects of extra dimensions on $g_{\mu}-2$ 
are substantially smaller than the effects from supersymmetric
loop corrections\cite{Nath:1999aa}. Experimentally the situation is still somewhat
unsettled due to the ambiguity in the errors of the hadronic
corrections. A recent estimate of the difference between 
experiment and the standard  model gives\cite{hagiwara}  
a $\sim 3 \sigma$ effect. 
Further, the sign of the difference in these analyses is positive
indicating a positive $\mu$. In Fig.1 we give a  numerical analysis
of neutralino-proton cross section $\sigma_{\chi p}$ which enters
in the direct detection of dark matter. The results are taken from
the analysis of Ref.\cite{Chattopadhyay:2002jx}. The analysis shows that the 
 $g_{\mu}-2$ constraint is very strong and eliminates a large part
 of the parameters space consistent with the relic density constraints.
 The analysis also shows that a very substantial part of the parameter
 space  consistent with $g_{\mu}-2$ and relic density constraints
 will be accessible to future dark matter detectors. 
 
We turn now to the implications of a positive $\mu$ for Yukawa coupling
unification and the implications of the Yukawa unification constraint
on supersymmetric dark matter analyses. It is known that 
$b-\tau$ unification prefers $\mu<0$\cite{Baer:2001yy}. The above arises
from the fact that a negative correction to the b-quark mass is
desired for $b-\tau$ unification and a negative correction is 
most readily manufactured if $\mu$ is negative. 
 However, a closer scrutiny shows that the sign of the loop correction
 to the b quark mass is not rigidly tied to the sign of $\mu$. 
 Thus the dominant correction to the b quark mass arises from the 
 gluino exchange diagram and the sign of this correction is determined
 by the sign of $\mu M_{\tilde g}$ (see 
 Ref.\cite{Ibrahim:2003ca} and the references quoted therein).
 At same time the supersymmetric correction to $g_{\mu}-2$ 
 is governed mainly by the chargino exchange and the sign
 of that contribution is determined by the term
 $\mu \tilde m_2$\cite{Ibrahim:1999aj}. 
 Thus it is possible to relax the rigid relationship between the
 $\mu$ sign and the sign of the loop correction to the b quark 
mass while maintaining the usual connection between the sign of 
$\mu$ and the sign of the supersymmetric loop correction to 
 $g_{\mu}-2$. The solution to this relaxation is provided by 
 nonuniversalities which allow one to switch the sign of 
 $M_{\tilde g}$ relative to the sign of $\tilde m_2$. 
This switch in sign can be seen to arise group theoretically 
when the gaugino mass terms has nontrivial group transformations.
Thus, for example, in $SU(5)$ the gaugino masses in general 
 tranform like the $(24\times 24)_{s}$ representation of $SU(5)$.
 Now $(24\times 24)_{s}=1+24+75+200$ and for the $24$ plet on the
 right hand side one has that the $SU(3)\times SU(2)\times U(1)$ 
 gaugino masses are in the ratio $M_3:M_2:M_1 = 2:-3:-1$ and
 there is a relative sign between the $M_3$ and the $M_2$. 
 Similarly for $SO(10)$ the gaugino
  masses transform like the $(45\times 45)_{sym}$ representations of
 $SO(10)$ where $(45\times 45)_{sym}=1+54+210+770$.
 Here the $54$ plet representation  on the right hand  gives\cite{nonuniso10}
  $M_3:M_2:M_1 = 1:-3/2:-1$ and one finds once again that 
 $M_3$ and $M_2$ have opposite signs.    
The above  possibilities allow
 for\cite{Chattopadhyay:2001va,Chattopadhyay:2001mj} (see also\cite{komine})
$\mu>0, ~\Delta a_{\mu}  >0,~\Delta_b^{\tilde g}<0$ allowing
for $b-\tau$ unification for a positive $\mu$. Detailed analyses of 
these can be found in\cite{Chattopadhyay:2001va,Chattopadhyay:2001mj}
 where the implications
of Yukawa unification on supersymmetric dark matter are also 
discussed.  

\begin{figure}           
\vspace*{-1.0in}                                 
\subfigure[]{                       
\label{xamutan5a} 
\hspace*{-0.6in}                     
\begin{minipage}[b]{0.5\textwidth}                       
\centering
\includegraphics[width=\textwidth,height=\textwidth]{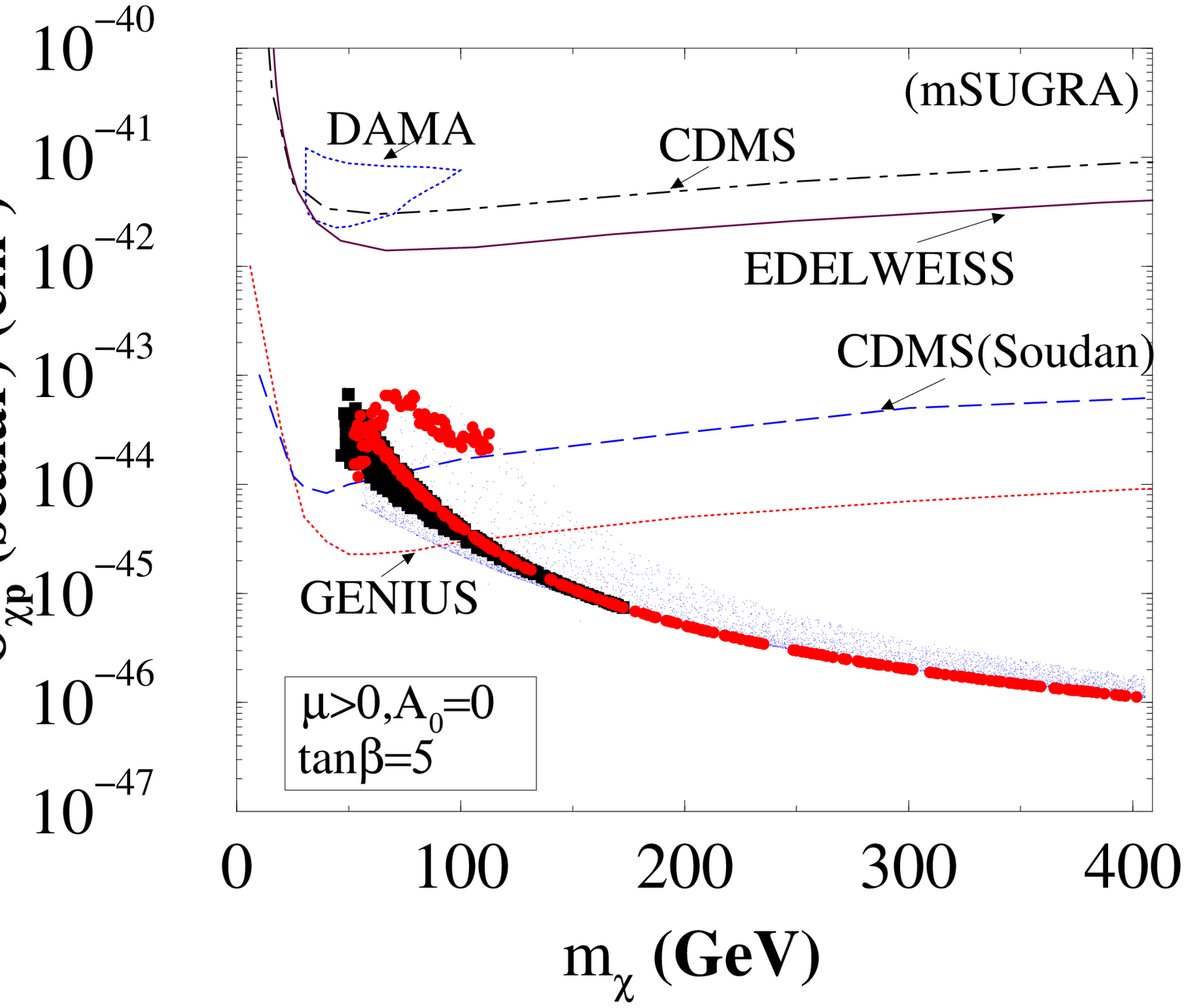}    
\end{minipage}}                       
\hspace*{0.3in}
\subfigure[]{      
\label{xamutan10}                  
\begin{minipage}[b]{0.5\textwidth}                       
\centering                      
\includegraphics[width=\textwidth,height=\textwidth]{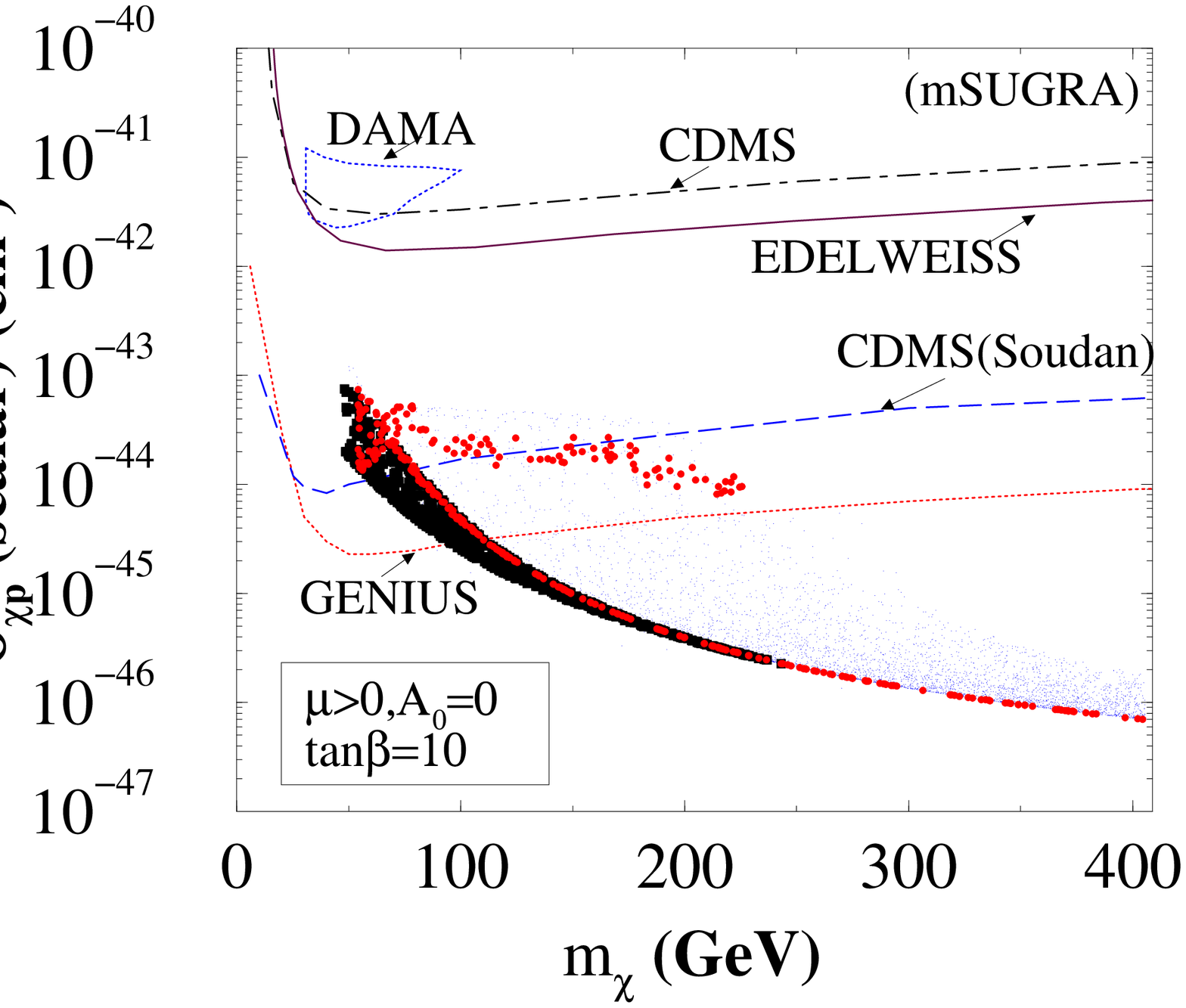} 
\end{minipage}}                       
\hspace*{-0.6in}                     
\subfigure[]{                       
\label{xamutan30}                  
\begin{minipage}[b]{0.5\textwidth}                       
\centering
\includegraphics[width=\textwidth,height=\textwidth]{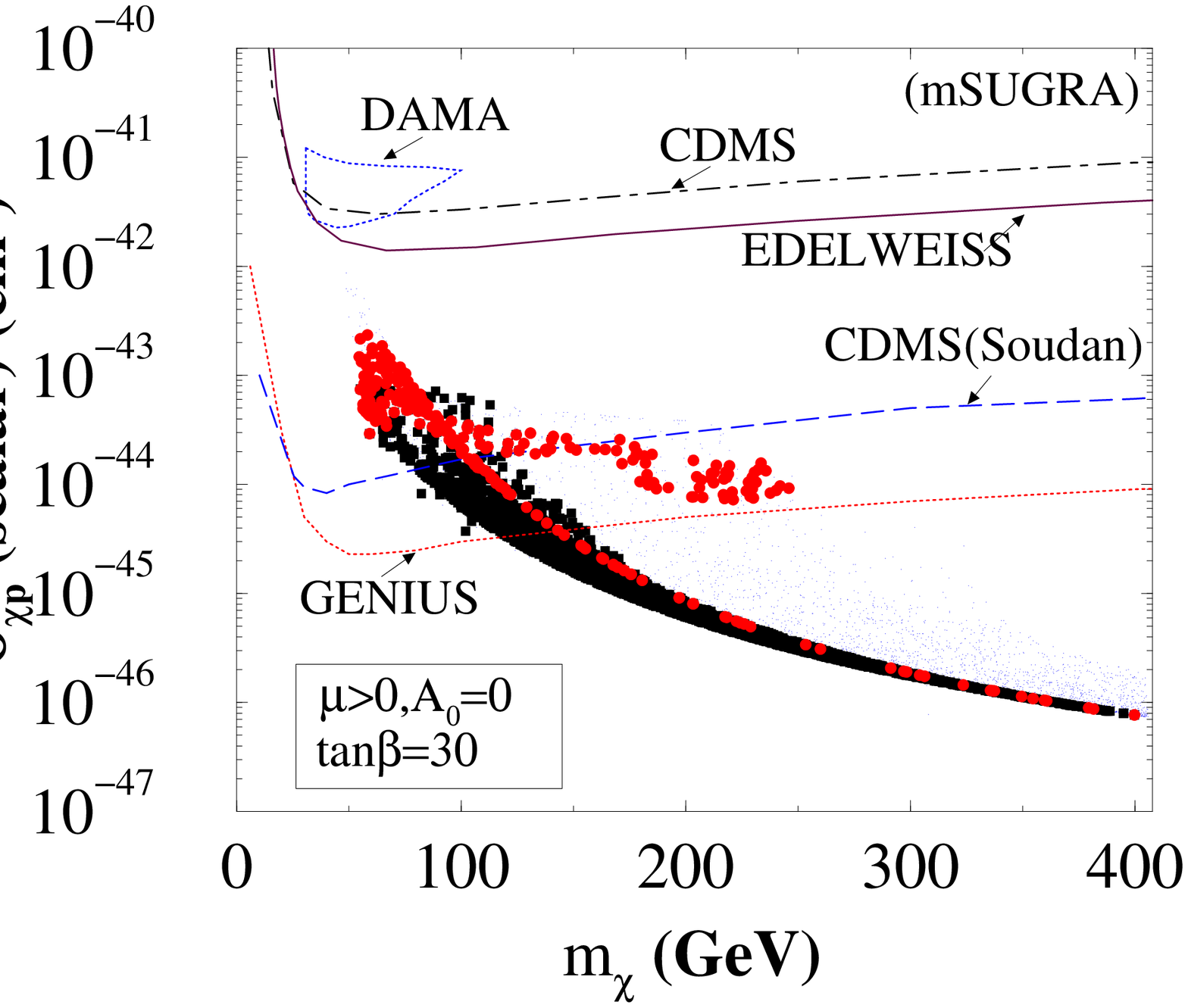}
\end{minipage}}
\hspace*{0.3in}                       
\subfigure[]{                       
\label{xamutan45a}
\begin{minipage}[b]{0.5\textwidth}                       
\centering                      
\includegraphics[width=\textwidth,height=\textwidth]{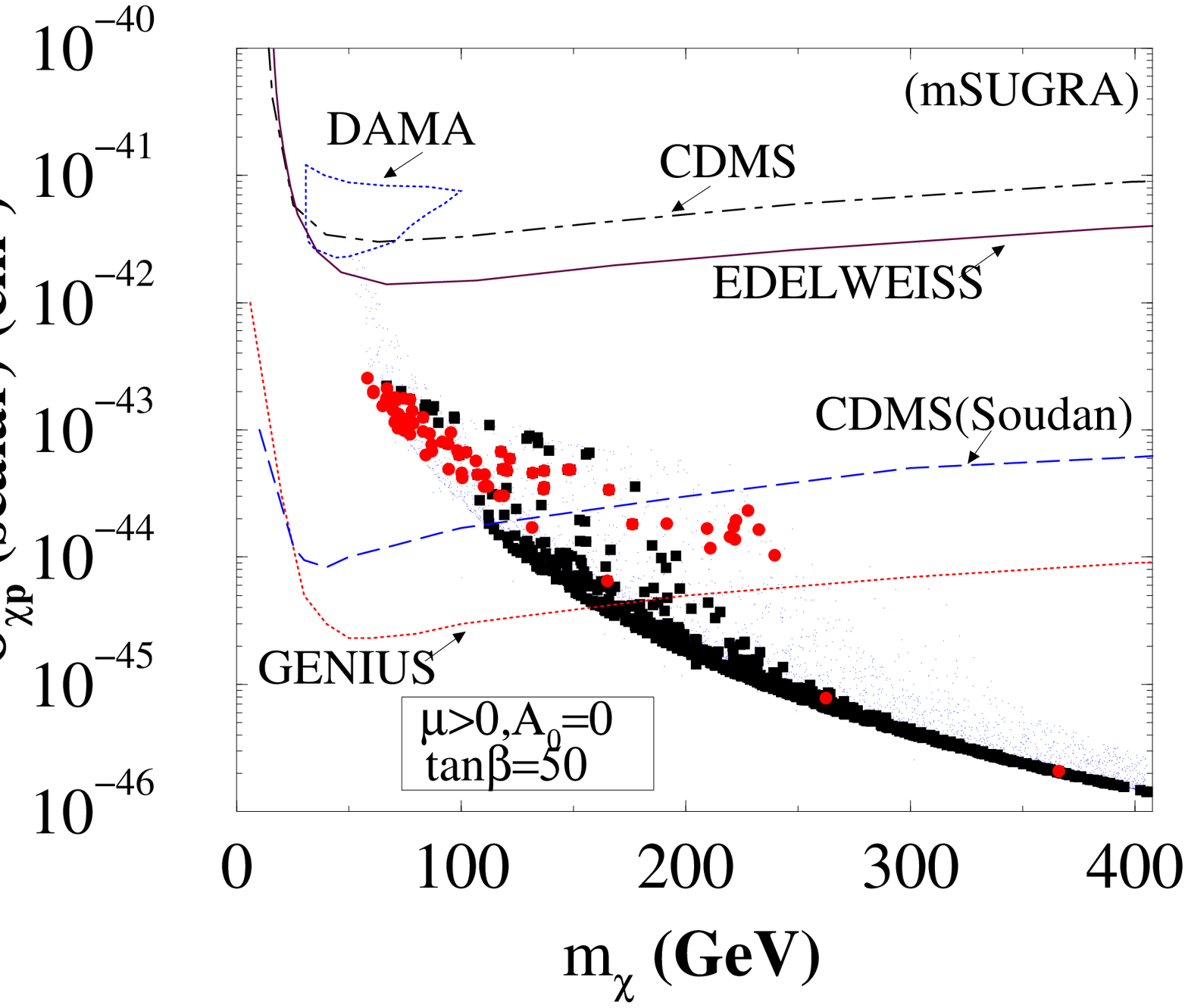}
\end{minipage}}
\caption{Exhibition of the neutralino-proton scalar cross section
for values of $\tan\beta$ of 5,10,30,50 and $A_0=0$ and $\mu>0$.
The blue dots are the points satisfying mSUGRA constraints and the
red circles  additionally satisfy the
 relic density constraint of $0.1<\Omega h^2<0.3$.
The black squares satisfy the $2\sigma$ constraint on 
$b\rightarrow s+\gamma$ and $g_{\mu}-2$. The limits  of various
current and future experiments are also indicated. 
Taken from Ref.\cite{Chattopadhyay:2002jx}}.
\label{xamutana} 
\end{figure}
\begin{figure}
\hspace*{-0.6in}
\centering
\includegraphics[width=8cm,height=8cm]{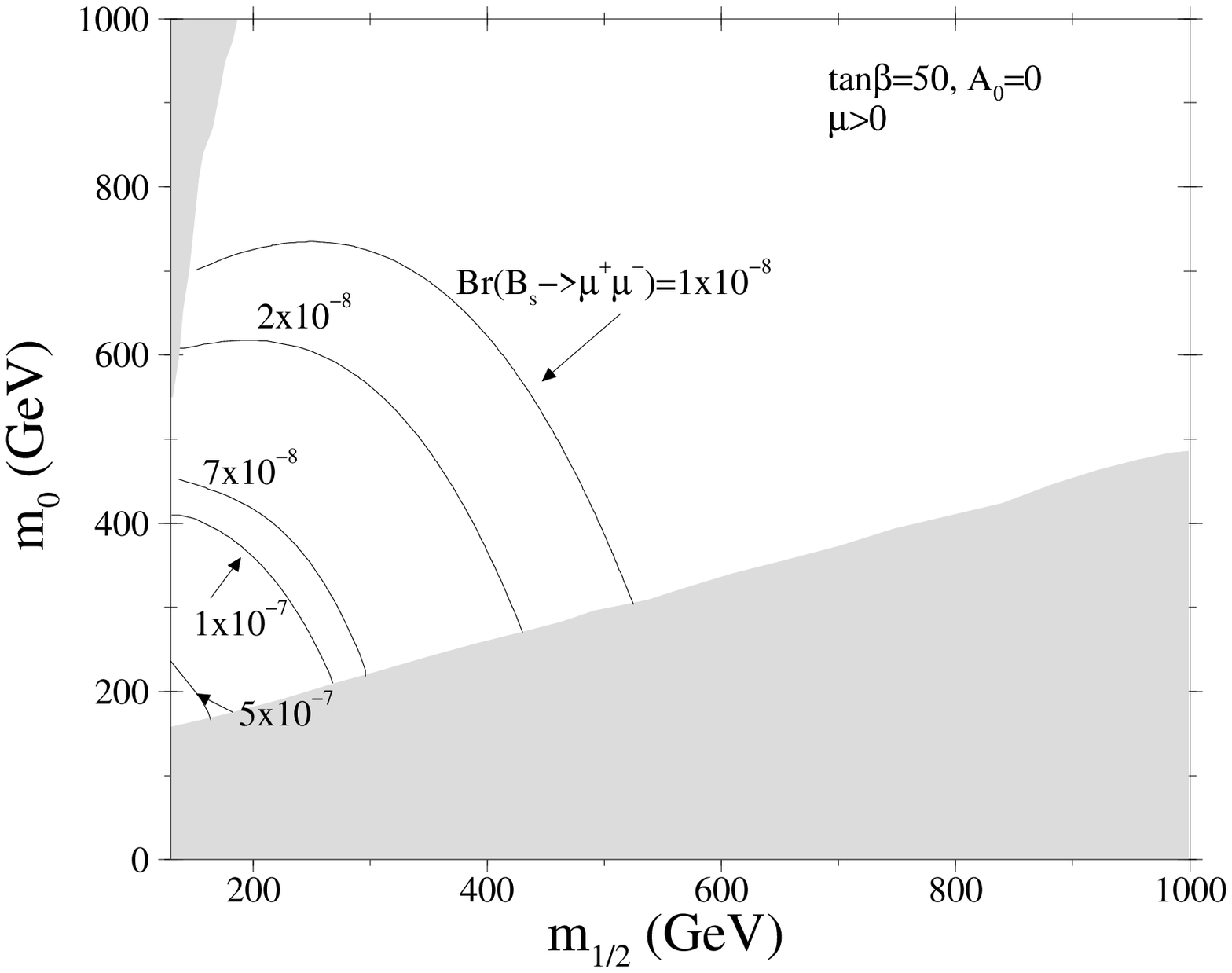}
\caption{An exhibition of the allowed parameter space in $m_0-m_{1/2}$ 
 from the limits  on $B^0_s\rightarrow \mu^+\mu^-$ branching ratio.
 Taken from Ref.\cite{Chattopadhyay:2003xi}} 
\label{bmumu}
\end{figure}

Next we discuss the implications of the  $B^0_{s,d}\rightarrow l^+ l^-$ 
constraint. In the Standard Model
 $ B(\bar B^0_s\rightarrow \mu^+\mu^-)=(3.1\pm 1.4)\times 10^{-9} $.
However, this branching ratio may lie beyond the reach of RUNII which may
reach a sensitivity of $10^{-(7-8)}$. It turns out that in SUSY/SUGRA 
models the $ B(\bar B^0_s\rightarrow \mu^+\mu^-)\sim \tan^6\beta$ 
for large $\tan\beta$\cite{gaur}. This arises from the so called 
counterterm diagram. Because of the large  $\tan\beta$ factor
the branching ratio in SUSY models can get much larger than in the 
Standard Model. Specifically in mSUGRA the branching ratio can 
be  as large as $O(10^{-6})$ and within reach of RUNII\cite{dedes}.
Further, it was found  that CP phases can provide an 
extra enhancement of $\sim 10^2$ in some cases (see Ibrahim and Nath in
Ref.\cite{dedes}).
Thus SUSY enhancement brightens the prospect for the observation of
this decay at Fermilab. Further, 
the observation of $B^0_{s,d}\rightarrow l^+ l^-$ will be evidence 
for SUSY  even before the sparticles are seen directly at colliders.
In Fig.~(\ref{bmumu}) the implications of the constraints  of various
 limits  on  $B^0_{s,d}\rightarrow l^+ l^-$  branching ratio
are given. Thus, for example, a branching ratio 
$BR(B^0_{s,d}\rightarrow l^+ l^-)=10^{-8}$ can probe the parameter  space 
in $m_{\frac{1}{2}}-m_0$ in the range up ot $600$ GeV-$700$ GeV for
$\tan\beta = 50$, $A_0=$ and $\mu>0$.  

\section{The Hyperbolic Branch, Supersymmetry and Dark Matter}
It was shown quite sometime ago that the radiative breaking of supersymmetry
has two branches: an ellipsoidal branch (EB) and a hyperbolic branch (HB).
This can be exhibited  easily by examining the relation for radiative breaking
that determines  $\mu$\cite{Chan:1997bi}, i.e., 
\beqn
C_1m_0^2+C_3m'^2_{1/2}+C_2'A_0^2+\Delta \mu^2_{loop}=
	\mu^2+\frac{1}{2}M_Z^2
\eeqn
where $\Delta \mu^2_{loop}$ arise from loop corrections to the 
effective potential\cite{Arnowitt:qp}.
When $\tan\beta$ is small, the loop corrections are typically small, and 
furthermore  $C_1,C_2',C_3>0$ and the variation of these parameters with the
renormalization group scale are also small. In this case the radiative
breaking of the electroweak symmetry is realized on the ellipsoidal branch
and there is an upper limit to the soft parameters for a fixed value of $\mu$.
However, for other regions  of  the parameter space specifically when $\tan\beta$
is large one  finds  that $\Delta \mu^2_{loop}$ is large and furthermore
the scale dependence of some of the co-efficients $C_i$ is rather large.
Specifically in this case, if one chooses a scale $Q_0$ where the loop
correction $\Delta \mu^2_{loop}$ is minimized one finds that  the
co-efficient $C_1$ at $Q_0$ turns negative  and the hyperbolic branch is realized.
Thus on the hyperbolic branch  $m_0, m_{\frac{1}{2}}$ can get very large 
for fixed $\mu$. A very interesting phenomenon of inversion
occurs  when $M_i>>|\mu|$. In this
case an examination of the neutralino mass matrix and of the chargino  mass matrix
shows that  $\chi_1^0, \chi_2^0$, $\chi_1^{\pm}$ are essentially 
degenerate with mass $\mu$.  More specifically one finds 
that\cite{Chattopadhyay:2003xi} (for other analyses that explore
the implications of WMAP data see \cite{elliswmap,hb/fp})
\beqn
M_{\chi_1^0}= \mu -\frac{M_Z^2}{2} (1-\sin 2\beta) 
[\frac{\sin^2\theta_W}{M_1-\mu} + \frac{\cos^2\theta_W}{M_2-\mu}]\nonumber\\
M_{\chi_2^0}= \mu +\frac{M_Z^2}{2} (1+\sin 2\beta) 
[\frac{\sin^2\theta_W}{M_1+\mu} + \frac{\cos^2\theta_W}{M_2+\mu}]\nonumber\\
M_{\chi_1^{\pm}}= \mu +\frac{M_W^2 \cos^2 \beta}{\mu}
-\frac{M_W^2}{\mu} \frac{(M_2\cos\beta + \mu \sin\beta)^2}{(M_2^2-\mu^2)}
\eeqn
In the inversion region the squarks and sleptons and the gluino may lie 
in the several TeV region and thus may not be 
easily observable at accelerators.
Here the lightest particles are 
$h^0, \chi_1^0, \chi_2^0, \chi_1^{\pm}$ where 
$m_{\chi_1^0}\simeq m_{\chi_1^{\pm}}\simeq m_{\chi_2^{0}}\simeq \mu$.
We note that the mass relations here are in gross violation of the
scaling laws\cite{scaling}.
Further, the quantities relevant for the observation of the lightest
supersymmetric particles in this case  
are the mass differences $\Delta M^{\pm} = m_{\chi_1^{\pm}}- m_{\chi_1^0}$,
and $\Delta M^{0} = m_{\chi_2^{0}}- m_{\chi_1^0}$. 
While $m_{\chi_1^0}$, $m_{\chi_1^{\pm}}$, $m_{\chi_2^{0}}$ may 
lie in the several hundred GeV region,
 $\Delta M^{\pm}$ and $\Delta M^{0}$ lies in the 1-10 GeV region. 
 The smallness of the mass differences makes 
the observation of these particles 
 rather difficult since the decay of the NLSP and also of the light chargino 
 will result in rather soft particles which may not be observable 
 at the LHC. Situations of this type have  been discussed
 in other contexts in the literature\cite{cdg,fmrss}.
  However, quite surprisingly the 
satisfaction of the relic density constraints can occur easily in 
the inversion region.
This is so because as pointed out already that in the inversion region 
there is a near degeneracy of $m_{\chi_1^0}$, $m_{\chi_1^{\pm}}$, 
$m_{\chi_2^{0}}$ which implies a lot of coannihilation.
Specifically one has coannihilations\cite{Edsjo:1997bg} involving the 
processes\cite{Chattopadhyay:2003xi}
\beqn
\chi_1^{+} \chi_1^{-}, 
\chi_1^0 \chi_2^{0}\rightarrow u_i\bar u_i, d_i \bar d_i, W^+W^-\nonumber\\
\chi_1^0 \chi_1^{+}, \chi_2^0 \chi_1^{+}\rightarrow u_i\bar d_i, \bar e_i\nu_i, AW^+,Z W^+,
W^+h 
\eeqn 
Because of coannihilation relic density constraints can be easily 
satisfied even though the squark and slepton masses may lie in the several
TeV region. Detailed analyses  show that in the HB region  
 $\sigma_{\chi_1^0-p}$ may still be accessible to dark matter experiment.
Thus observation of dark matter may be the only means of 
observing SUSY effects if the inversion region of HB  is realized
in nature. We note in passing that the so called focus point 
region\cite{fmm} is a part of the hyperbolic branch and corresponds to
low values of $m_{\frac{1}{2}}$. For further discussion 
see Ref.\cite{Chan:1997bi,Chattopadhyay:2003xi,elliswmap,hb/fp}.

\begin{figure}
\hspace*{-0.6in}
\centering
\includegraphics[width=8cm,height=8cm]{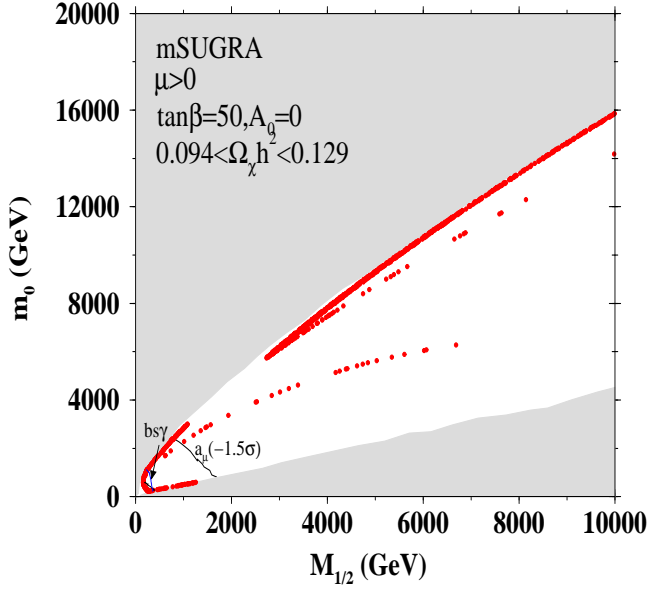}
\caption{Regions allowed by the WMAP relic density constraints 
in the $m_0-m_{1/2}$  plane for 
$\tan\beta =50$ with a $2\sigma$ error corridor around the 
 WMAP constraint. Taken from Ref.\cite{Chattopadhyay:2003xi}} 
\label{m0mhalf}
\end{figure}

\begin{figure}           
\vspace*{-1.0in}                                 
\subfigure[]{                       
\label{xamutan5} 
\hspace*{-0.6in}                     
\begin{minipage}[b]{0.5\textwidth}                       
\centering
\includegraphics[width=\textwidth,height=\textwidth]{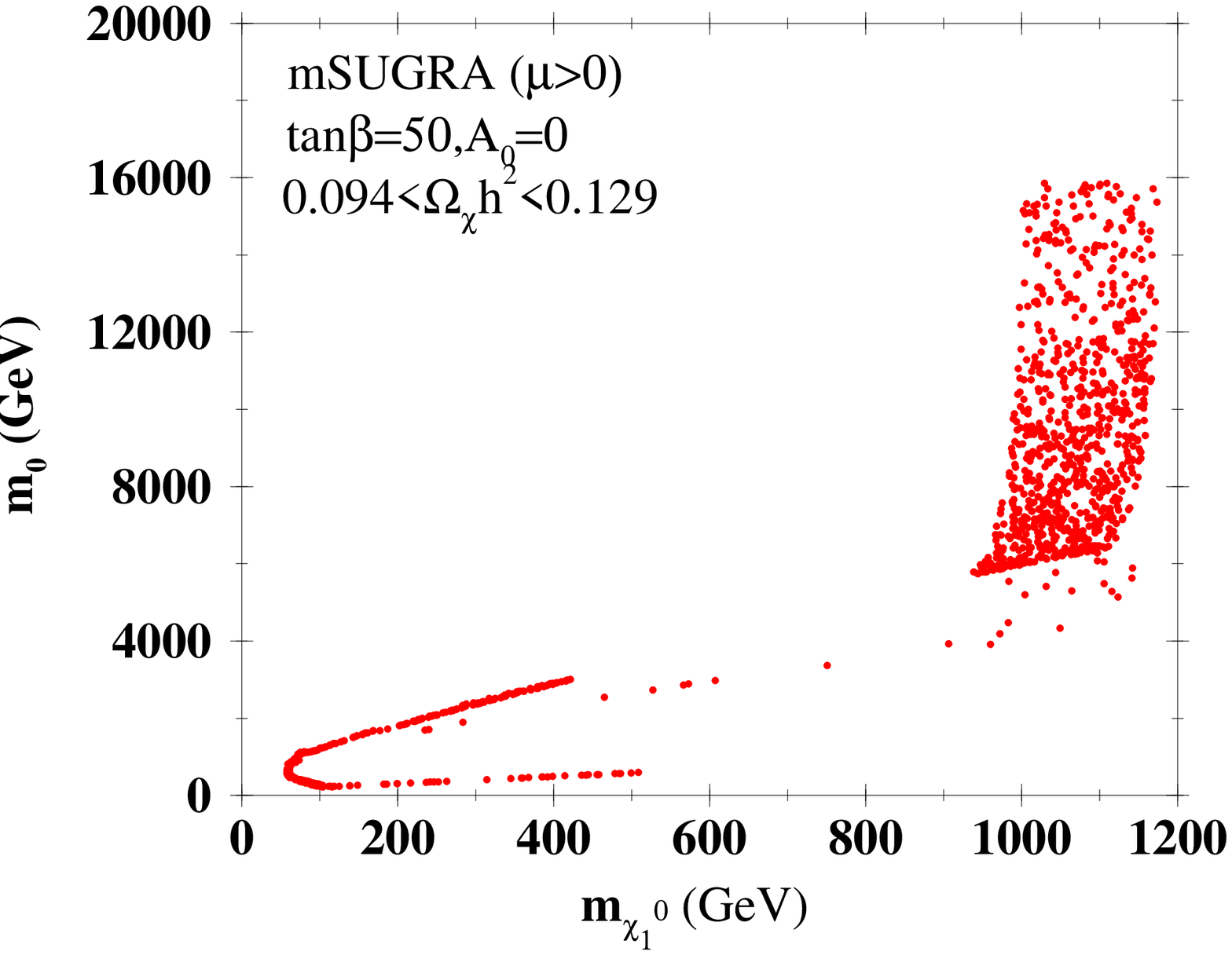}   
\end{minipage}}                       
\hspace*{0.3in}
\subfigure[]{      
\label{xamutan10a}                  
\begin{minipage}[b]{0.5\textwidth}                       
\centering                      
\includegraphics[width=\textwidth,height=\textwidth]{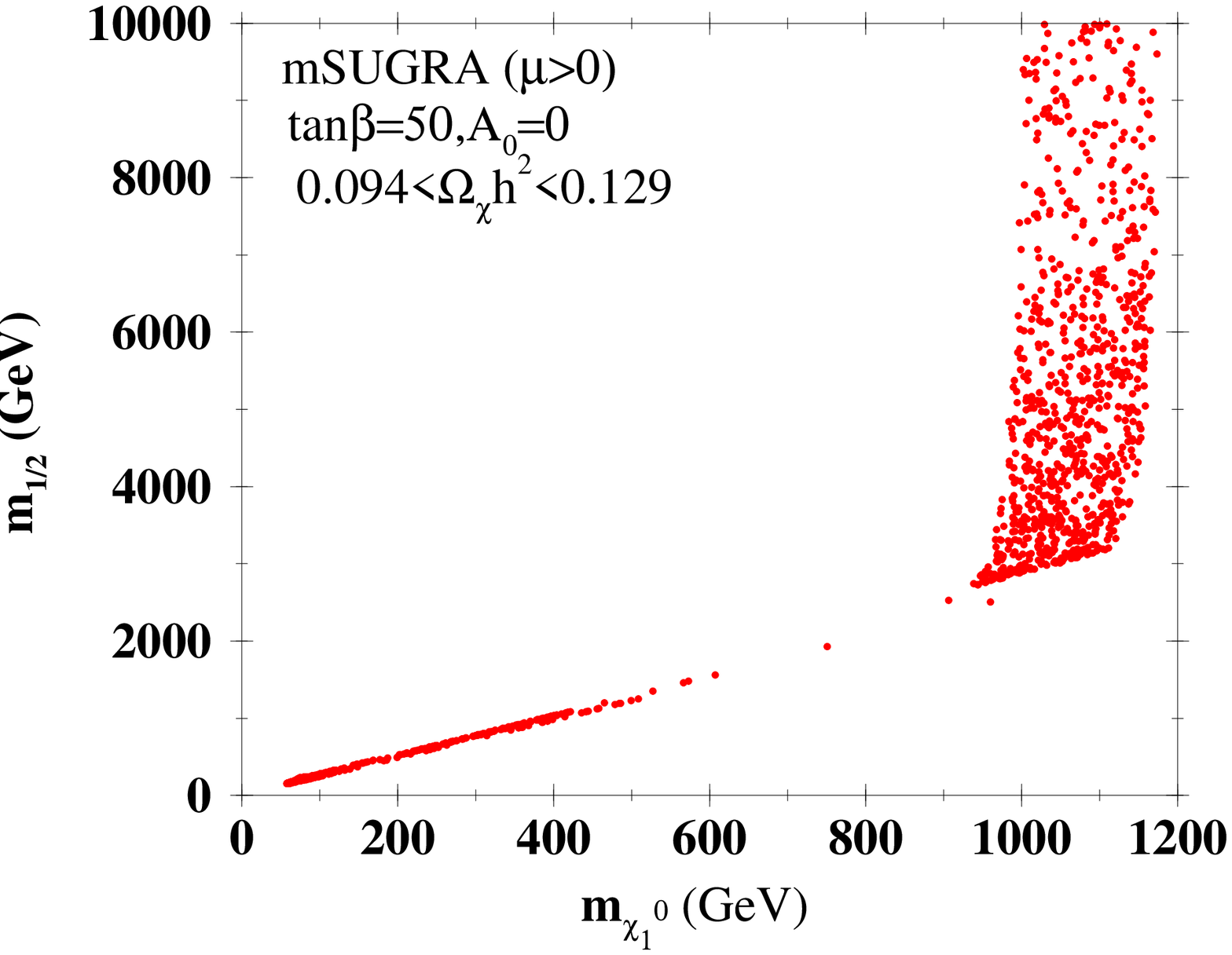} 
\end{minipage}}                       
\hspace*{-0.6in}                     
\subfigure[]{                       
\label{xamutan30a}                  
\begin{minipage}[b]{0.5\textwidth}                       
\centering
\includegraphics[width=\textwidth,height=\textwidth]{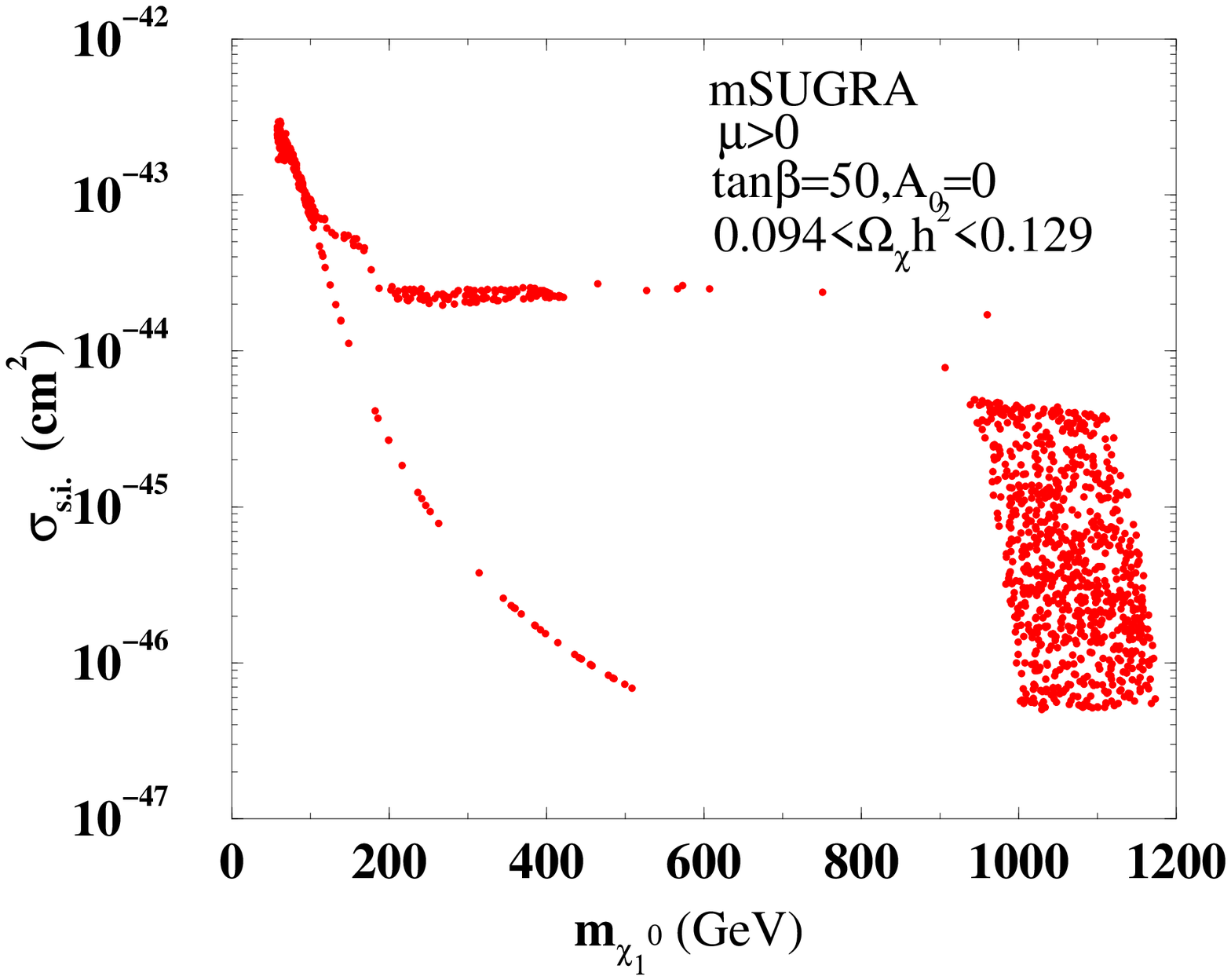} 
\end{minipage}}
\hspace*{0.3in}                       
\subfigure[]{                       
\label{xamutan45}
\begin{minipage}[b]{0.5\textwidth}                       
\centering                      
\includegraphics[width=\textwidth,height=\textwidth]{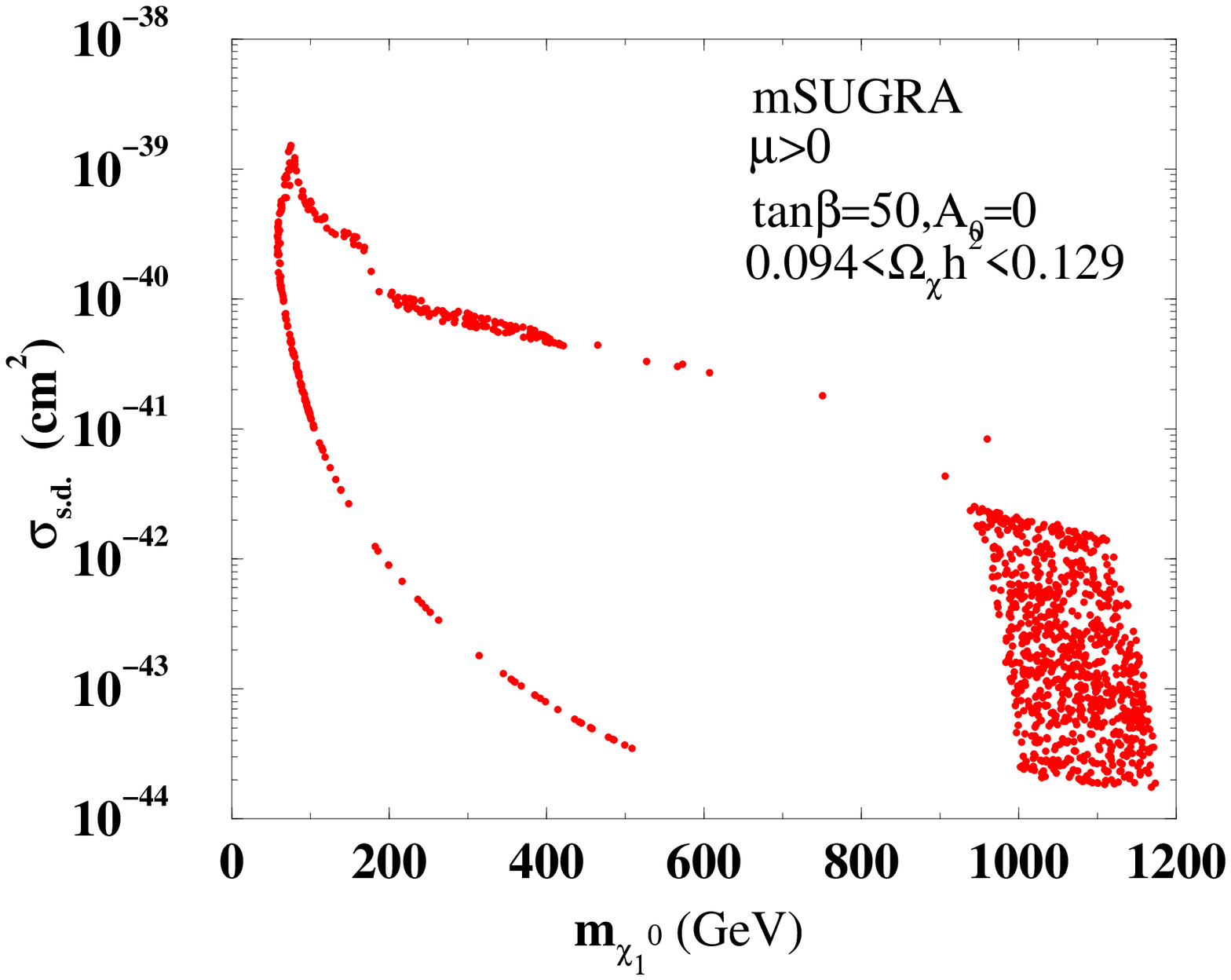} 
\end{minipage}}
\caption{(a) Plot in $m_0-m_{\chi_1^0}$ showing the allowed  region
consistent with WMAP constraints; (b) Same  as  (a) except the plot is
in $m_0-m_{\frac{1}{2}}$; (c) Scalar $\sigma_{\chi p}$ cross section
as a  function of $m_{\chi_1^0}$; (d) Same  as (c) except that the plot is
for spin dependent $\sigma_{\chi p}$ cross section. The patches  
exhibit the inversion region where the neutralino is  essentially
a higgsino. Taken from Ref.\cite{Chattopadhyay:2003xi}.}
\label{xamutan} 
\end{figure}
\section{Analysis  of dark matter with WMAP constraints}
In this section we  discuss the implications of the WMAP
 constraints. 
   We also discuss the implications for the
 detection of dark matter in dark matter detectors presently
 in operation\cite{dama,cdms,hdms,edelweiss} as well as those
 that are planned for the future\cite{genius,cline}. 
 We begin with the result of WMAP which gives for CDM\cite{bennett,spergel}
\beq
\Omega_{\chi}h^2 = 0.1126^{+0.016}_{-0.018}
\label{wmapeqn}
\eeq
 In the theoretical analyses of the relic density 
 coannihilation\cite{mizuta,efo1,efo2,gomez,ads,efo3,nihei,bednyk}
  plays  a central
 role. However, unlike the usual coannihilation phenomena where the
 particles that enter in the coannihilation  process are the neutralino 
 and the stau here the particles that coannihilate are 
 $\chi_1^0, \chi_2^0.\chi_1^{\pm}$. Specifically the processes are listed
 in Eq.(3). In the inversion region the neutralino is essentially a
 Higgsino as opposed to being a Bino which is what happens over the most 
 of the rest of the parameter space of SUGRA models. 
 In Fig.(3) a numerical analysis of the region in the parameter space
of $m_0-m_{\frac{1}{2}}$ allowed under the WMAP relic density constraint
with a $2\sigma$ error corridor is exhibited. One finds that 
the region consistent with the WMAP relic density constraint can
indeed stretch to rather large values in ($m_0$, $m_{\frac{1}{2}}$)
extending into several TeV in each direction. In Figs.(4) a further 
investigation of the parameter space consistent with the relic density
constraints is carried out. Thus in Fig.(4a) an analysis of the parameter
space in the $m_0$ and $m_{\chi_1^0}$ consistent with the relic
density constraint is given. Here one finds that $m_{\chi_1^0}$
is sharply limited  while $m_0$ gets large. The vertical patch 
at $m_{\chi_1^0}$ around a TeV is the inversion region where the
neutralino is essentially a Higgsino. A similar phenomenon is visible in
Fig.(4b). A plot of the neutralino-proton scalar (spin
dependent) cross section $\sigma_{\chi p}$ as a function of
 $m_{\chi_1^0}$ is given in 
Fig.(4c)(Fig.(4d)). A comparison with Fig.(1) shows that a 
significant part of the inversion region will be accessible to 
future experiments on the direct detection of dark matter. 

\section{Conclusion}
In this paper we have given a brief summary of some of the
recent developments in supersymmetric dark matter. We have
discussed the constraints of $g_{\mu}-2$ and of 
$B^0_s\rightarrow \mu^+\mu^-$ on dark matter analyses. 
One of the most stringent constraints arises from the recent
observation from WMAP which has measured the relic density for
CDM to a high degree of accuracy. We discussed the allowed
parameter space in mSUGRA satisfying the WMAP constraints.
It was shown that quite surprisingly the allowed parameter
space is quite large.
 Specifically one finds that a very significant
region on the hyperbolic branch with ($m_0$, $m_{\frac{1}{2}}$)
extending in several TeV still allows the satisfaction
of the relic density constraints consistent with WMAP. The consistency 
with the WMAP data arises due to coannihilation of 
$\chi_1^0,\chi_2^0,\chi_1^{\pm}$. Further, one finds 
that the neutralino-proton cross section fall in range that 
may be accessible to dark matter detectors in the future.
Thus if SUSY is realized deep on the hyperbolic branch, then 
direct observation of sparticles, aside from the light Higgs, may 
 be difficult. However, degeneracy of $\chi_1^0$, 
$\chi_2^0$, $\chi_1^{\pm}$ would lead to significant coannihilation
and satisfaction of relic density constraints and the direct
detection of supersymmetric dark matter may still be possible.
Finally, we note that in heterotic string models $\tan\beta$ is a 
determined quantity under the constraints of radiative breaking
of the electroweak symmetry\cite{Nath:2002nb} and thus dark matter
analyses are more constrained in this framework. This constraint
will be explored in further work.
A similar situation may occur in models based on soft breaking in
intersecting D branes\cite{Kors:2003wf}.
\\

  \noindent
 {\bf Acknowledgments}\\ 
  This research was supported in part by NSF grant  PHY-0139967

\end{document}